\documentclass[preprint,prc,showpacs,preprintnumbers,
                unsortedaddress,amsmath,amssymb,floatfix]{revtex4}
\usepackage{graphicx}
\usepackage{dcolumn}
\usepackage{bm}
\usepackage{longtable}
\usepackage{color}
\usepackage[bookmarksnumbered,plainpages]{hyperref}
\newcommand{\ben}{\begin{enumerate}}
\newcommand{\een}{\end{enumerate}}
\newcommand{\beq}{\begin{equation}}
\newcommand{\eeq}{\end{equation}}
\newcommand{\beqn}{\begin{eqnarray}}
\newcommand{\eeqn}{\end{eqnarray}}

\newcommand{\beqd}{\begin{eqnarray*}}
\newcommand{\eeqd}{\end{eqnarray*}}
\newcommand{\bea}{\begin{array}}
\newcommand{\eea}{\end{array}}
\newcommand{\bcen}{\begin{center}}
\newcommand{\ecen}{\end{center}}
\newcommand{\btab}{\begin{tabular}}
\newcommand{\etab}{\end{tabular}}
\newcommand{\bsub}{\begin{subequations}}
\newcommand{\esub}{\end{subequations}}
\newcommand{\bit}{\begin{itemize}}
\newcommand{\eit}{\end{itemize}}
\newcommand{\brule}{\begin{ruledtabular}}
\newcommand{\erule}{\end{ruledtabular}}
\newcommand{\bpm}{\begin{pmatrix}}
\newcommand{\epm}{\end{pmatrix}}

\newcommand{\cals}[1]{{\cal #1}}
\newcommand{\tr}{{\rm Tr}}

\newcommand{\ff}[1]{\frac{1}{#1}}

\newcommand{\lb}{\left(}
\newcommand{\rb}{\right)}
\newcommand{\ls}{\left[}
\newcommand{\rs}{\right]}

\newcommand{\svec}[1]{{\mbox{\boldmath${\rm #1}$}}}
\newcommand{\ivec}{\vec}

\newcommand{\bay}{\begin{array}}
\newcommand{\eay}{\end{array}}

\newcommand{\balp}{\mbox{\boldmath$\alpha$}}

\newcommand{\bp}{{\mathbf{p}}}

\newcommand{\br}{{\mathbf{r}}}

\newcommand{\bP}{{\mathbf{P}}}

\begin{document}
 \title{Effects of the triaxial deformation and pairing correlation on the
       proton emitter $^{145}$Tm}
 \author{J. M. Yao}
 \affiliation{School of Physics, Peking University, 100871 Beijing, China}
 \author{B. Sun}
 \affiliation{School of Physics, Peking University, 100871 Beijing, China\\
              Gesellschaft f\"ur Schwerionenforschung GSI, 64291 Darmstadt,
              Germany}%
 \author{P. J. Woods}
 \affiliation{University of Edinburgh, Edinburgh, EH9 3JZ, United
 Kingdom}
 \author{J. Meng}
 \email[]{mengj@pku.edu.cn}
 \affiliation{School of Physics, Peking University, 100871 Beijing, China\\
              Department of Physics, University of Stellenbosch, Stellenbosch, South
              Africa \\
              Institute of Theoretical Physics, Chinese Academy of
              Sciences, Beijing, China\\
              Center of Theoretical Nuclear Physics,
              National Laboratory of Heavy Ion Accelerator, 730000 Lanzhou, China}%
 \date{\today}
 \begin{abstract}%
 \vspace{2em}
  The ground-state properties of the recent reported proton emitter $^{145}$Tm
  have been studied within the axially or triaxially deformed relativistic mean
  field (RMF) approaches, in which the pairing correlation is taken into account by the
  BCS-method with a constant pairing gap. It is found that triaxiality and
  pairing correlations play important roles in reproducing the experimental
  one proton separation energy. The single-particle level, the proton emission orbit,
  the deformation parameters $\beta=0.22$ and $\gamma=28.98^\circ$ and the corresponding
  spectroscopic factor for $^{145}$Tm in the triaxial RMF calculation are
  given as well.
 \end{abstract}
 \maketitle
 \section{Introduction}\label{Introduction}

 Proton-rich nuclei display many interesting structural properties which
  are important both for nuclear physics and astrophysics. These nuclei are
  characterized by exotic decay modes, such as the direct emission of charged
  particles from the ground state, and $\beta$-decays with large Q-values. The
  decay via direct proton emission provides a unique insight into the structure
  of nuclei beyond the drip line limit. The evolution of the single-particle
  structure, nuclear shapes and masses can be deduced from measured properties
  of proton emission~\cite{Woods97}.

  Proton emission in both spherical and deformed systems has been studied extensively
  in the past decades~\cite{Delion06}. For the spherical proton emitter, a simple WKB
  estimation of the transmission through the Coulomb and centrifugal barriers could give
  the correct order of magnitude of the decay rates and the angular momentum of the decaying
  state ~\cite{Feix83,Iriondo86,Davids97,Aberg97}. Most of the proton emitters in the rare earth
  region, which are predicted to have large static quadrupole deformations~\cite{Moeller95}
  are analyzed by a particle-coupled core model with the unbound proton interacting with an
  axially symmetric deformed core~\cite{Kadmensky96,Ferreira97,Davids98,Maglione98,Maglione99,Sonzogni99,Rykaczewski99,Barmore00,Esbensen00,Seweryniak01}.
  Such an analysis over the past several years turned out to be a good description of the ground-state
  properties of axially deformed rare-earth proton emitters.

  Recently, the proton emission from triaxial nuclei has drawn lots of attention~\cite{Kruppa04,Davids04,Delion04,Arumugam07,Seweryniak07}.
  Specific combinations of single-particle orbitals near the Fermi surface can
  lead to a propensity toward triaxial shapes as illustrated by recent calculations
  of the additional binding energy due to non-axial degrees of freedom~\cite{Moeller06},
  which revealed several "islands" of triaxiality throughout the nuclear chart.
  In Ref.~\cite{Davids04}, the static triaxial deformation was introduced
  in the adiabatic coupled channels method in order to investigate proton
  emission from $^{141}$Ho($7/2^-$). The total decay rate and the 2+ branching
  ratio, however, were found to be good agreement with experimental data only for
  the triaxial angle $\gamma<5^\circ$. The importance of triaxial deformation
  in proton emitters $^{161}$Re and $^{185}$Bi was pointed out in Ref.~\cite{Delion04},
  where the decay widths were found to be very sensitive to the $\gamma$ deformation.
  However, the sensitivity of decay widths in $^{161}$Re and $^{185}$Bi
  on the triaxial deformation was questioned in a recent paper reporting a non-adiabatic quasipartilce
  calculation~\cite{Arumugam07}. Instead, the pairing effect was found to have a more significant
  influence. In Ref.~\cite{Seweryniak07}, the quasiparticle-coupled core model has been used to
  address the important role of the gamma degree of freedom in the prediction of the proton
  decay rate and the spectrum of excited states in the proton emitter $^{145}$Tm.

  It should be pointed out that the nuclear potentials in the (quasi)particle-coupled
  core model are tuned to fit the measured energy of the decaying state~\cite{Esbensen00}.
  In particular, it cannot provide any information about the microscopic structure properties of proton
  emitters.

  Approaches based on concepts of non-renormalizable effective relativistic
  field theories and density functional theory provide a very reliable theoretical framework for
  studies of nuclear structure phenomena at and far from the valley of $\beta$-stability. In particular,
  the relativistic mean field (RMF) theory, which can take into account the spin-orbit coupling
  naturally, has been successfully applied in analysis of nuclear structure over the whole periodic
  table, from light to superheavy nuclei with a few universal parameters~\cite{Serot86,Reinhard89,Ring96,Vretenar05,Meng06}.
  The application of RMF theory with the restriction of axial symmetry to study the properties of
  proton emitter has already been done~\cite{Vretenar99,Lalazissis99,Geng04}. In general,
  the predicted location of the proton drip line,
  the ground-state quadrupole deformations, one proton separation
  energies at and beyond the drip line, the deformed single-particle orbit occupied by the
  odd valence proton, and the corresponding spectroscopic factor are in good agreement with the
  experimental data. However, the influence of triaxiality on proton emitters has not been
  investigated in the microscopic self-consistent RMF approach. Here in this paper, the
  influences of $\gamma$ deformation degree of freedom and pairing correlations on
  proton emitters $^{145}$Tm will be studied within the
  triaxial deformed RMF approach.

  The paper is arranged as follows. In Sec.~\ref{model},  a brief introduction of
  the RMF approach will be given.
  Both axial and triaxial calculations with pairing correlation are carried out to
  investigate the properties of proton emitter $^{145}$Tm in Sec.~\ref{Results},
  including total energy, quadrupole deformations, one-proton separation energy,
  the potential of the valence proton, and the corresponding spectroscopic factor.
  Finally, our conclusions and summary are given in Sec.~\ref{Summary}.

 \section{The relativistic mean field theory}\label{model}
 The starting point of the RMF theory with meson-exchange providing nucleon-nucleon
 interaction is the standard effective Lagrangian density constructed with the degrees
 of freedom associated with the nucleon field~($\psi$), two isoscalar meson fields
 ~($\sigma$ and $\omega_\mu$), the isovector meson field~($\vec\rho_\mu$) and the
 photon field~($A_\mu$),
 \beqn
 \label{Lagrangian}
 \begin{split}
 \cals L = ~&\bar\psi\ls i\gamma^\mu\partial_\mu - m
            - g_\sigma\sigma
            - g_\omega\gamma^\mu\omega_\mu
            - g_\rho\gamma^\mu\ivec\tau\cdot\ivec\rho_\mu
            - \ff2 e(1-\tau_3)\gamma^\mu A_\mu\rs\psi \\
          & + \ff2\partial^\mu\sigma\partial_\mu\sigma
                - U_\sigma(\sigma)
            -\ff4\Omega^{\mu\nu}\Omega_{\mu\nu}
                + U_\omega(\omega_\mu)
            -\ff4 \ivec R^{\mu\nu}\cdot\ivec R_{\mu\nu}
                + U_\rho(\ivec\rho_\mu) \\
          & -\ff4 F^{\mu\nu}F_{\mu\nu},
 \end{split}
 \eeqn
where $m$ and $m_i(g_i)$ ($i = \sigma,\omega_\mu,\vec\rho_\mu$) are
the masses (coupling constants) of the nucleon and the mesons
respectively and
 \bsub
 \beqn
\Omega^{\mu\nu}  &=& \partial^\mu\omega^\nu
                   - \partial^\nu\omega^\mu, \\
\ivec R^{\mu\nu} &=& \partial^\mu\ivec\rho^\nu
                   - \partial^\nu\ivec\rho^\mu, \\
   F^{\mu\nu}    &=& \partial^\mu A^\nu
                   - \partial^\nu A^\mu
 \eeqn
 \esub
are the field tensors of the vector mesons and the electromagnetic
field. Here in this paper, we adopt the arrows to indicate vectors
in isospin space and bold types for the space vectors. Greek indices
$\mu$ and $\nu$ run over 0, 1, 2, 3, while Roman indices $i$, $j$,
etc. denote the spatial components.

The nonlinear self-coupling terms $U_\sigma(\sigma)$,
$U_\omega(\omega_\mu)$, and $U_\rho(\ivec\rho_\mu)$ for the
$\sigma$-meson, $\omega$-meson, and $\rho$-meson in the Lagrangian
density (\ref{Lagrangian}) respectively have the following forms:
 \bsub
 \beqn
 U_\sigma(\sigma)     &=& \ff2m_\sigma^2\sigma^2
                        + \ff3 g_2\sigma^3 + \ff4 g_3\sigma^4, \\
 U_\omega(\omega_\mu) &=& \ff2m_\omega^2\omega^\mu\omega_\mu
                        + \ff4 c_3\lb\omega^\mu\omega_\mu\rb^2, \\
 U_\rho(\ivec\rho_\mu)&=& \ff2m_\rho^2\ivec\rho~^\mu\cdot\ivec\rho_\mu.
 \eeqn
 \esub
In the mean-field approximation, the correspondent energy functional
is obtained as
 \beq
  \label{EnergyF}
  \begin{split}
   E_{\text{RMF}}[\rho,\phi_m]
    = & \int d^3x \tr\left[ \beta\left( \svec\gamma\cdot\svec p + m + g_\sigma\sigma
        + g_\omega\gamma_\mu\omega^\mu  + g_\rho\gamma^\mu\ivec\tau\cdot\ivec\rho_\mu
        + \ff2 e(1-\tau_3)\gamma^\mu A_\mu\right)\rho\right]\\
      & + \int d^3x\left\{\ff2\partial^0\sigma\partial_0\sigma
        - \ff2\partial^i\sigma\partial_i\sigma + U_\sigma(\sigma)
        - \ff4\Omega^{0\mu}\Omega_{0\mu}
        + \ff4\Omega^{i\mu}\Omega_{i\mu} - U_\omega(\omega_\mu)\right.\\
     &~~~~~~~~~~~~~~\left.
        - \ff4\ivec R^{0\mu}\cdot\ivec R_{0\mu}
        + \ff4\ivec R^{i\mu}\cdot\ivec R_{i\mu}
        - U_\rho(\ivec\rho_\mu)
        - \ff4 F^{0\mu}F_{0\mu} + \ff4 F^{i\mu}F_{i\mu}\right\},
 \end{split}
 \eeq
where $\phi_m$ denotes $\{\sigma, \omega_\mu, \ivec\rho_\mu, A_\mu\}$ respectively.

The equations of motion for the nucleon and the mesons can be
obtained by requiring that the energy functional (\ref{EnergyF}) be
stationary with respect to the variations of $\rho$ and $\phi_m$,
 \beq
  \label{stationary}
  \delta \left\{ E_{\text{RMF}}[\rho, \phi_m] -\tr(\epsilon \rho) \right\}= 0,
 \eeq
where $\epsilon$ is a diagonal matrix, whose diagonal elements are
the single particle energies. Using the variation $\delta\rho$ with
respect to $\psi_k$, the stationary condition (\ref{stationary})
leads to the Dirac equation,
 \beq
  \label{DiracEq}
  [\balp\cdot\bp+\beta(m + S +\gamma^\mu V_\mu)]\psi_k = \epsilon_k\psi_k
 \eeq
for the nucleon. The scalar potential $S$ and vector potential $V_\mu$ in Eq. (\ref{DiracEq})
are respectively,
 \bsub
 \beqn
 \label{potential}
 S     &=& g_\sigma\sigma, \label{potential-scalar}\\
 V_\mu &=& g_\omega\omega_\mu + g_\rho\vec\tau\cdot\vec\rho_\mu
             + \ff2 e(1-\tau_3)A_\mu \label{potential-vector}.
 \eeqn
 \esub
 The Klein-Gordon equations for the mesons and the photon are given by,
 \beq
 \label{MesonEqs}
  \partial_\mu\partial^\mu \phi_m + U^\prime(\phi_m)
  =\pm S_{\phi_m},
 \eeq
 where the $(+)$ sign is for vector fields and the $(-)$ sign for
 the scalar field. The source terms $S_{\phi_m}$ in Eq.(\ref{MesonEqs})
 are sums of bilinear products of Dirac spinors
 \beqn
   S_{\phi_m}=
  \left\{\begin{array}{ll}
         \sum\limits_{k>0}v^2_k\overline{\psi}_k \psi_k, & \phi_m=\sigma \\
         \sum\limits_{k>0}v^2_k\overline{\psi}_k \gamma_\mu\psi_k, & \phi_m=\omega_\mu\\
         \sum\limits_{k>0}v^2_k\overline{\psi}_k \gamma_\mu \vec\tau\psi_k, & \phi_m=\vec\rho_\mu\\
         \sum\limits_{k>0}v^2_k\overline{\psi}_k \gamma_\mu\frac{1-\tau_3}{2}\psi_k,& \phi_m=A_\mu
    \end{array}
  \right.
 \eeqn
 where the sums run over only the positive-energy states ($k>0$) (i.e., no sea approximation) and
 the occupation probability of the single-particle energy level $k$, i.e., $v^2_k$, is evaluated
 within the BCS method. It is sufficient for proton-rich nuclei because of the attenuating effect
 of the Coulomb barrier on the spatial extent of the proton wave function, which limits the formation
 of proton halo.

 \section{Results and discussion}\label{Results}

\begin{figure}[h!]
 \centering
  \includegraphics[height=6cm]{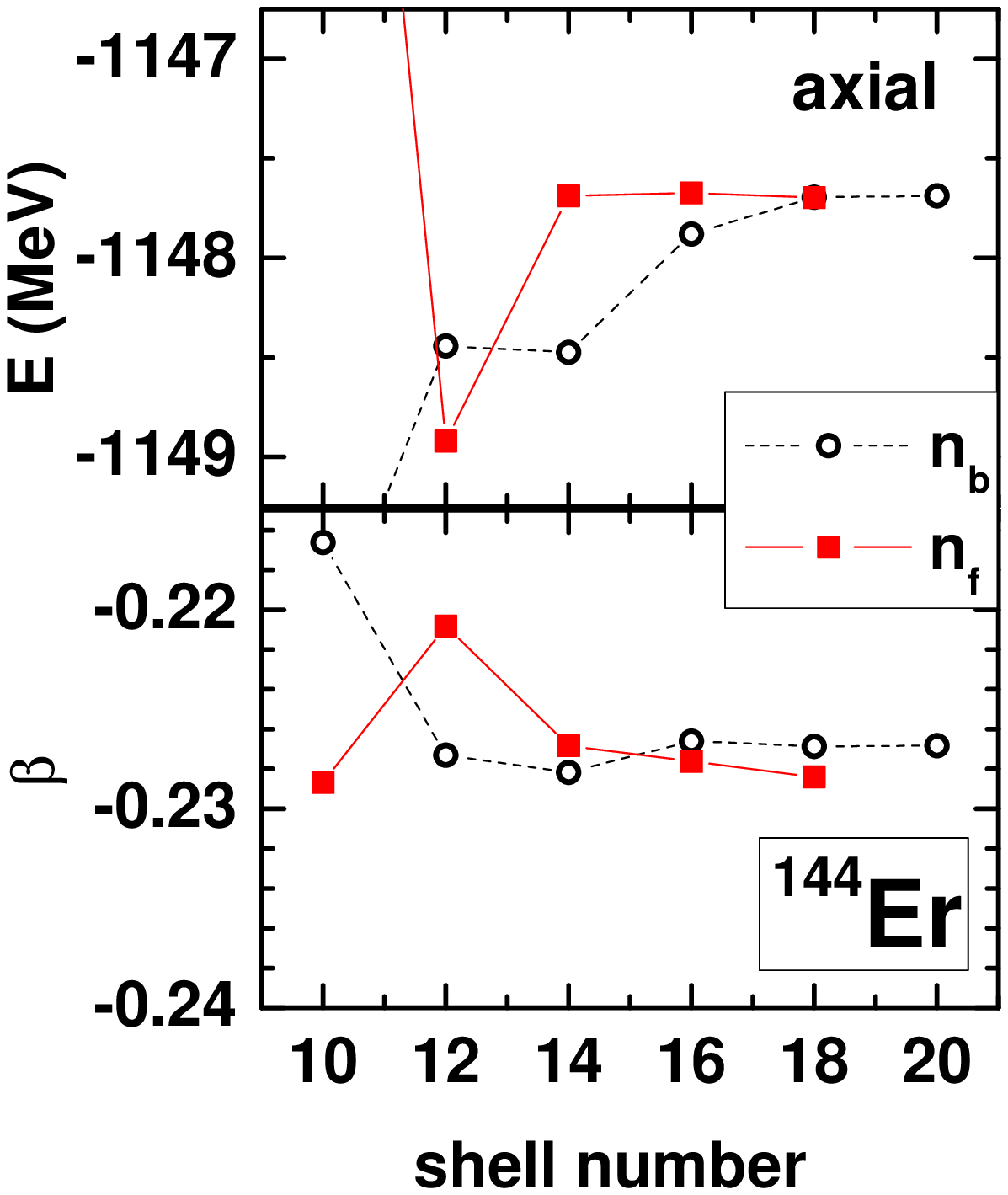}
  \includegraphics[height=6cm]{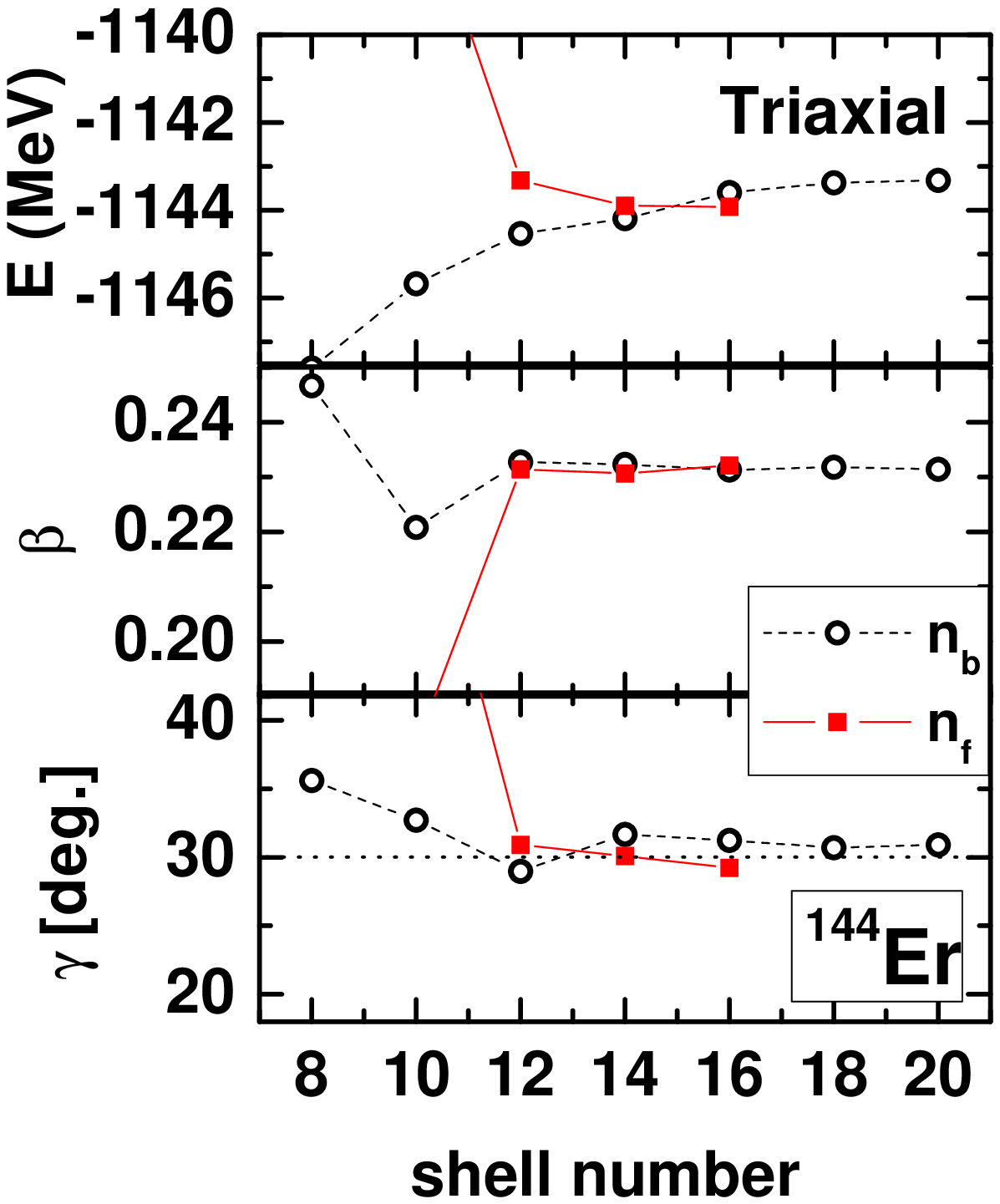}
   \caption{(Color online) The total energy, deformation parameters $\beta$ and $\gamma$
   calculated in axially (left panel) and triaxially (right panel) deformed RMF with
   PK1 for $^{144}$Er as functions of the number of expanded oscillator shells for the meson field
   $n_{\rm b}$ and the nucleon field $n_{\rm f}$.}
   \label{fig1}
 \end{figure}

 The static Dirac equation (\ref{DiracEq}) for the nucleon and Klein-Gordon
 equation (\ref{MesonEqs}) for the meson fields are solved by expansion on the cylindric
 (axially deformed RMF) or three-dimensional Cartesian (triaxially deformed RMF) harmonic
 oscillator basis with major shell numbers as $n_{\rm f}$ and $n_{\rm b}$ for the nucleons
 and mesons respectively.  The equation of motion for the
 photon field is solved using the standard Green's function method because of its long range.
 The parameter set PK1~\cite{Long04} is used throughout the
 calculation and the center-of-mass (c.m.) correction is taken into
 account by
 \beqn
  \label{Eq:Ecm}
   E^{\rm mic}_{\rm cm}
   = \frac{1}{2mA}\langle\hat \bP^{2}_{\rm cm}\rangle,
 \eeqn
 where $\bP_{\rm c.m.}$ is the total momentum of a nucleus with $A$ nucleons.
 In order to check the convergence of the results with the number
 of expanded oscillator shells for nucleons $n_{\rm f}$ and for mesons $n_{\rm b}$, the
 total energy, quadrupole deformation $\beta$ and $\gamma$ of $^{144}$Er as functions
 of shell numbers are calculated with axially (left panel) and triaxially (right panel) deformed
 RMF approaches as shown in Fig.~\ref{fig1}. It indicates that as long as $n_{\rm f}\geq 14,
 n_{\rm b}\geq 18$, the binding energies and the deformations are independent of the
 expanded shell numbers. Therefore in the following, $n_{\rm f}=14$ and $n_{\rm b}=20$ will be adopted.
 More details about the axially and triaxially deformed RMF approaches can be found in
 Ref.~\cite{Gambhir90} and Refs.~\cite{Hirata96,Yao06} respectively.

 In order to get the pairing gaps for $^{144}$Er and $^{145}$Tm, we fit odd-even mass
 differences of around 120 nuclides~\cite{Audi03} in the very proton-rich side ranging
 from La to Re by the four-point difference formula.
 The neutron and proton pairing gaps obtained are, $\Delta^{\rm est.}_n=13.7/\sqrt{A}$ and $\Delta^{\rm est.}_p=15.9/\sqrt{A}$
 respectively, and the corresponding rms deviation with respect to experimentally known values
 are 0.17 and 0.16 MeV. Considering the blocking effect of odd valence proton, the proton pairing
 gap for $^{145}$Tm is reduced by a factor $f$, ranging from zero to one in
 the calculation. For the odd-mass system, the blocking calculations are performed
 without breaking the time-reversal invariance. In this case, the space-like components
 of the vector fields vanish and this may introduce uncertainty around
 several hundred keV~\cite{Yao06}, which may be compensated by the pairing gap.

\begin{figure}[h!]
 \centering
  \includegraphics[width=6cm]{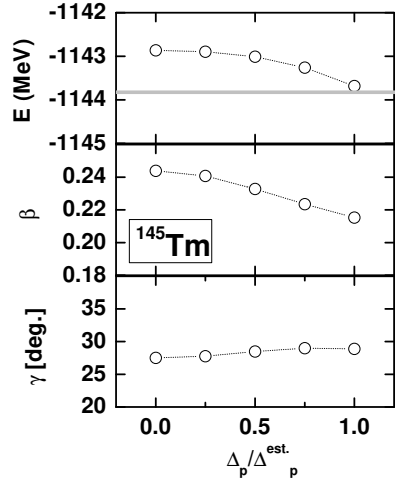}
   \caption{The total energy, deformation parameters $\beta$ and $\gamma$ as functions
            of the proton pairing reduction factor in
            $^{145}$Tm calculated by the triaxial RMF approach with PK1, where a reduction
            factor $f=\Delta_p/\Delta^{\rm est.}_p$ varying from 0 to 1 has
            been used in the BCS approximation and the data
            is given by a grey line.}
   \label{fig2}
 \end{figure}

 The total energy, deformation  parameters $\beta$ and $\gamma$ in $^{145}$Tm
 as functions of the pairing reduction factor $f$ are investigated in the triaxial RMF+BCS/PK1
 approach and plotted in Fig.~\ref{fig2}, in which the total experimental energy for $^{145}$Tm
 is obtained from the systematic estimated atomic binding energy in Ref.~\cite{Audi03} by
 subtracting the electron binding energy according to Ref.~\cite{Lunney03}.

\begin{figure}[h!]
 \centering
  \includegraphics[width=6cm]{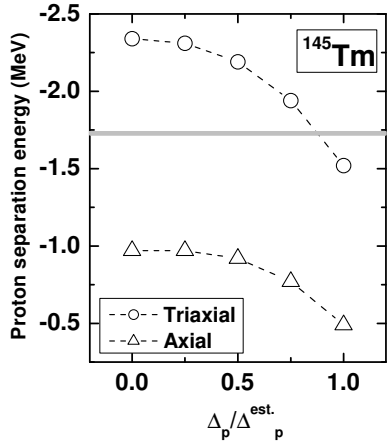}
   \caption{One proton separation energy plotted as a function of proton pairing reduction factor
            varying from 0 to 1 for $^{145}$Tm in the axial and triaxial RMF
            calculations with PK1, and pairing correlations treated by the BCS
            approximation. The neutron pairing gaps for  $^{145}$Tm and $^{144}$Er
            as well as the proton pairing gap for $^{144}$Er are chosen as the estimated values.
            The data is given by a grey line.}
   \label{fig3}
 \end{figure}

 It shows that the proton pairing gap $\Delta_p$ has
 a significant influence on the total energy, but a negligible influence on the
 predicted deformation parameters. Especially  the triaxility parameter $\gamma$
 is almost independent on $\Delta_p$. The one proton separation energy $S_p$ in $^{145}$Tm as a function
 of the $f$ is plotted in Fig.~\ref{fig3}. It is noticed that the $S_p$ in the triaxial RMF+BCS/PK1
 calculation coincides with the experimental value, $1.728(10)$~MeV~\cite{Batchelder98} with a
 certain $\Delta_p$ (i.e., $f\simeq0.9$). However, the axial RMF+BCS/PK1 calculation can not
 reproduce the experimental data of $S_p$. It indicates the importance of both the $\gamma$
 degree of deformation and pairing correlations in the description of proton emission in $^{145}$Tm.

 The one proton separation energy $S_p$, charge radius $r_c$, neutron radius $r_n$,
 as well as  deformation  parameters $\beta$ and $\gamma$, the single-particle orbital
 occupied by the odd valence proton, and the corresponding spectroscopic factor $u^2_k$
 for $^{145}$Tm are calculated in the axially deformed and triaxial RMF+BCS/PK1 approaches
 ( $f=0.9$ for proton in $^{145}$Tm).
 The spectroscopic factor $S_k$ of the deformed odd-proton orbital $k$ is
 approximately given by the unoccupied probability $u^2_{k}$ of state $k$ in the daughter
 nucleus with an even proton number~\cite{Lalazissis99}.
 The results are compared with those of relativistic Hartree-Bogoliubov (RHB),
 the Hartree-Fock-Bogoliubov (HFB-14), the finite-range droplet
 mass model (FRDM) as well as the experimental data in Table~\ref{Table1}.
 Similar to the HFB-14 prediction, the axially deformed RMF approach predicts an
 oblate shape with $\beta=-0.21$ for $^{145}$Tm, while the RHB shows a prolate ground-state.
 The different predicted shapes for $^{145}$Tm can be ascribed to the shape
 coexistence~\cite{Geng04}. While the large difference for one proton separation energy
 between the axially deformed RMF+BCS/PK1($S_p=0.62$~MeV) and the RHB
 calculation ($S_p=1.43$~MeV) is ascribed to the treatment of the pairing.

 After taking into account the $\gamma$ degree of deformation self-consistently,
 the one proton separation energy ($S_p=1.71$~MeV) in the triaxial
 RMF+BCS/PK1 calculation is in good agreement with the data.
 The corresponding deformation parameters
 are respectively $\beta=0.22$ and $\gamma=28.98^\circ$ as shown in
 Table~\ref{Table1}.
 Furthermore, the spectroscopic factor of the odd valence proton is 0.67, which is
 consistent with the value 0.51(16) obtained with the WKB approximation
 calculation~\cite{Batchelder98}.

\begin{figure}[h!]
 \centering
  \includegraphics[width=6cm]{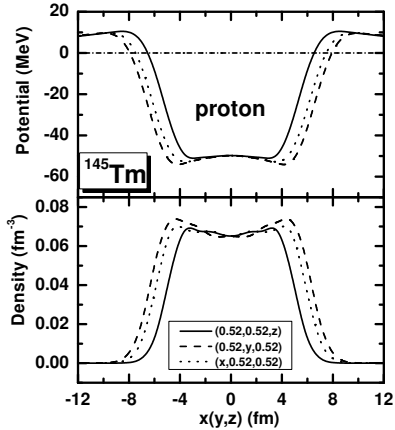}
   \caption{The mean-field potential and density distribution for   the
            protons plotted
            as functions of $x$ (for $y=0.52$~fm and $z=0.52$~fm) (dotted line),
            $y$ (for $x=0.52$~fm and $z=0.52$~fm) (dashed line), as well as $z$
            (for $x=0.52$~fm and $y=0.52$~fm) (solid line) in $^{145}$Tm.}
   \label{fig4}
 \end{figure}

 The triaxiality can make the proton tunneling through the
 Coulomb barrier easier. The mean-field potential and density distribution of protons
 in $^{145}$Tm are plotted in Fig.~\ref{fig4} as functions of $x$ (for $y=0.52$~fm
 and $z=0.52$~fm) (dotted line), $y$ (for $x=0.52$~fm and $z=0.52$~fm) (dashed line), as
 well as $z$ (for $x=0.52$~fm and $y=0.52$~fm) (solid line), respectively. It shows that both the
 potential and density distribution are triaxially deformed.
 The Coulomb barrier is different in different directions and obviously the Coulomb
 barrier in the direction $y$ is lower than those in $x$ and $z$.

\begin{figure}[h!]
 \centering
  \includegraphics[width=8cm]{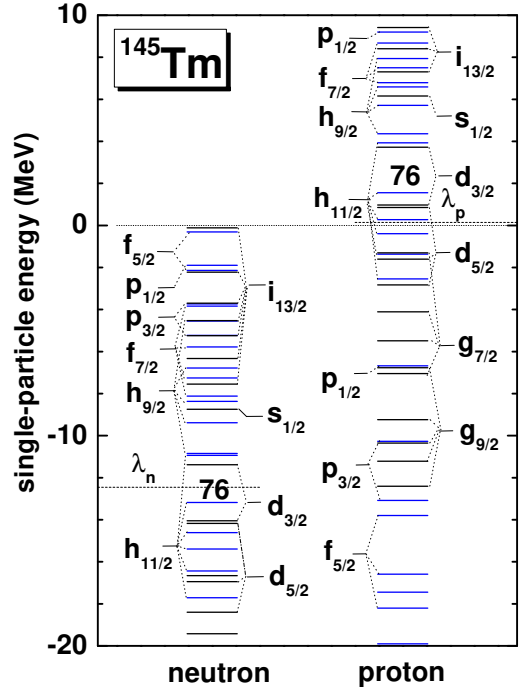}
   \caption{(Color online) Single-particle levels for neutrons and protons in $^{145}$Tm
            in the triaxial RMF calculation with PK1, and pairing correlations treated by the BCS
            approximation. The level is labeled with the quantum numbers of its main-component in the
            spherical Dirac spinor. The black (blue) one is the level with positive
            (negative) parity.}
 \label{fig5}
 \end{figure}

 Apart from the Coulomb barrier, the proton decay probability also depends on
 its energy and orbital angular momentum, i.e., the centrifugal barrier.
 In Fig.~\ref{fig5}, the single-particle energy levels for both neutron and proton
 in $^{145}$Tm are given, in which each level is labeled with the quantum numbers of
 its main-component in the spherical Dirac spinor. The valence proton
 belongs to the $h_{11/2}$ subshell, which is consistent with the observed spin-parity
 in Ref.~\cite{Seweryniak07}.

 In refs.~\cite{Rykaczewski01,Karny03},
 fine structure in proton emission was observed in $^{145}$Tm. In
 order to reproduce the experimental partial proton half-lives,
 it is found that the wave function of the valence proton in $^{145}$Tm
 is composed mainly of $67\%$ for $0h_{11/2}$ and $3.7\%$ for $1f_{7/2}$, which coupled
 to the ground state and the excited state of the $^{144}$Er core.
 Using the particle-core vibration model~\cite{Hagino01} and
 assuming deformation $\beta=0.18$, the experimental half-life and the fine
 structure branching ratio can be reproduced, in which
 the wave function of the valence proton
 is composed of $56\%$ for $0h_{11/2}$ and $3\%$ for $1f_{7/2}$~\cite{Karny03}.
 In contrast,  by reproducing the fine structure branching, the particle-core vibration
 calculations in ref.~\cite{Davids01} give $33\%$ only for $0h_{11/2}$.

\begin{figure}[h!]
 \centering
  \includegraphics[width=10cm]{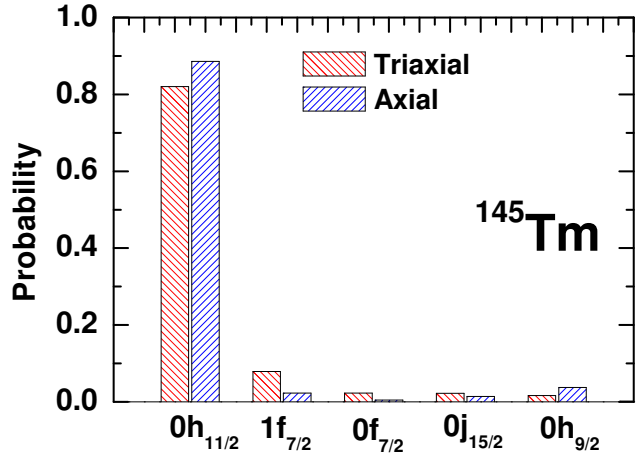}
   \caption{(Color online) Composition of the wave function for the valence
   proton in $^{145}$Tm calculated with axially and triaxially deformed RMF
   approaches. }
   \label{fig6}
 \end{figure}

 It is interesting to examine the composition of the valence proton in $^{145}$Tm
 obtained from the present microscopic and self-consistent RMF calculation.
 In Fig.~\ref{fig6}, the main spherical components of wave function for the valence proton
 in $^{145}$Tm by axially and triaxially deformed RMF calculations are given.
 The main components in axial RMF calculation
 are $88.6\%$ for $0h_{11/2}$, $3.7\%$ for $0h_{9/2}$ and
 $2.3\%$ for $1f_{7/2}$. While in triaxial RMF calculation, the
 main components are $82.1\%$ for $0h_{11/2}$,
 $7.9\%$ for $1f_{7/2}$, and $2.3\%$ for $0f_{7/2}$.

\begin{figure}[h!]
 \centering
  \includegraphics[width=6cm]{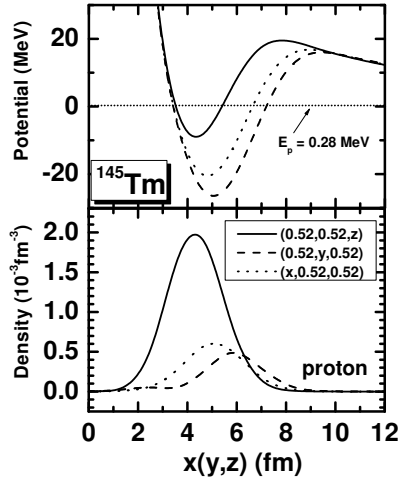}
   \caption{The proton-nucleus potential (upper panel) and density distribution of
                 the valence proton (lower panel) plotted as functions of
                 $x$ (for $y=0.52$~fm and $z=0.52$~fm) (dotted line),
                 $y$ (for $x=0.52$~fm and $z=0.52$~fm) (dashed line),
                 as well as $z$ (for $x=0.52$~fm and $y=0.52$~fm) (solid line)
                 in triaxial RMF calculation
                 with PK1 and pairing correlations treated by BCS
                 approximation. The proton Fermi level is
                 given by the short-dotted line (upper panel).}
 \label{fig7}
 \end{figure}

 The potential and density distribution for the valence proton are obtained
 in triaxial RMF calculation. To show what the triaxial potentials and
 proton distribution look like, we plot the potential and density
 distribution for the valence proton in Fig.~\ref{fig7}, in which the potential
 $V_{k}(\br)$ is given by,
  \beq
    V_{k}(\br) = V_0(\br) + S(\br) + \frac{(\hbar c)^2}{2m_pc^2}
                   [ \sum_{\mu} \vert F_{k\mu}\vert^2 V_\ell(r)
                   + \sum_{\mu^\prime}\vert G_{k\mu^\prime}\vert^2 V_{\ell^\prime}(r)],
    \quad V_\ell(r) = \frac{\ell(\ell+1)}{r^2},
  \eeq
  where $F_{k\mu}$ and $G_{k\mu^\prime}$ are respectively the expansion coefficients of the large
  and small components in spherical basis $\vert \mu \rangle=\vert n\ell
  jm_j\rangle$.

  A triaxial RMF+BCS/PK1 calculation with the estimated pairing gaps is also
  carried out for the neighboring proton emitters $^{146}$Tm and $^{147}$Tm.
  In contrast with the predicted shape transition from prolate to oblate
  existing from $^{145}$Tm to $^{146}$Tm~\cite{Moeller95,Lalazissis99},
  large triaxial deformations have been found for $^{146}$Tm ($\beta=0.19$, $\gamma=39.70^\circ$)
  and $^{147}$Tm ($\beta=0.21$, $\gamma=28.16^\circ$). The calculated one proton separation
  energies are $1.26$~MeV and $0.90$~MeV, which is close to the
  data, $1.120(10)$~MeV~\cite{Livingston93} and $1.054(19)$~MeV~\cite{Sellin93} respectively.

\section{Summary}\label{Summary}
 The axially and triaxially deformed RMF approaches have been applied
 for the description of ground-state properties of the proton emitter $^{145}$Tm
 with parameter sets PK1 and a constant pairing gap BCS-method for
 the pairing correlations. It has been found that the triaxiality and pairing correlations
 are essential to reproduce the one proton separation energy in $^{145}$Tm.
 The observed spin and parity of the emitted proton can be understood
 from the main spherical component by the transformation of the quantum number of the
 valence proton. The corresponding spectroscopic factor in $^{145}$Tm obtained
 in the present calculation is
 consistent with that obtained with the WKB approximation
 calculation. Large triaxial deformations have been found in
 $^{146}$Tm and $^{147}$Tm as well.

\begin{acknowledgments}

JM would like to thank University of Edinburgh for the hospitality
where this work is initialized and also SUPA distinguished programm
for the financial support. This work is partly supported by Major
State Basic Research Developing Program 2007CB815000 as well as the
National Natural Science Foundation of China under Grant No.
10435010, 10775004 and 10221003.

\end{acknowledgments}


\newpage

 \begin{table}[h!]
 \centering
 \tabcolsep=6pt
 \caption{Ground-state properties of the proton emitter $^{145}$Tm.
         The one proton separation energy $S_p$,
          charge radius $r_c=\sqrt{r^2_p+0.64}$, neutron $rms$ radius $r_n$, deformation
          parameters $\beta$ and $\gamma$, as well as the quantum numbers
          of its main-component in the spherical Dirac spinor $j^\pi$
          and spectroscopic factor $u^2_k$ for the valence proton
          in triaxial RMF calculation with PK1, and pairing correlations treated
          by the
          BCS approximation
          are compared with those of the axial RMF calculation with PK1, and pairing correlations treated
          by the
          BCS approximation (Axial), the relativistic Hartree-Bogoliubov  calculation (RHB),
          the Hartree-Fock-Bogoliubov calculation (HFB-14), the finite-range droplet mass model (FRDM) and
          experimental data. The neutron and proton pairing gaps in
          $^{144}$Er and $^{145}$Tm are estimated by $\Delta^{\rm est.}_n=13.7/\sqrt{A}$ MeV,
          $\Delta^{\rm est.}_p=15.9/\sqrt{A}$ MeV while the proton pairing gap in $^{145}$Tm
          is given by $0.9\times15.9/\sqrt{A}$ MeV. }
 \begin{tabular}{cccccccc}
 \hline
                                 & S$_p$ (MeV) & $r_c$ (fm) & $r_n$ (fm) & $\beta$  & $\gamma$     & $j \pi$ & $u^2_k$\\
  \hline
  Exp.~\cite{Batchelder98}       & -1.728       &            &            &          &              & $ 11/2-$          & 0.51(16) \\
  Triaxial                       & -1.71        &   5.078    &  4.995     &   0.22   & 28.98$^\circ$& $ 11/2-$  & 0.67 \\
  Axial                          & -0.62        &   5.073    &  4.992     &  -0.21   &              & $ 7/2-$[523]      & 0.53 \\
  RHB~\cite{Lalazissis99}        & -1.43        &            &            &   0.23   &              & $ 7/2-$[523]      & 0.47 \\
  HFB-14~\cite{Goriely07}        & -1.43        &   5.073    &            &  -0.20   &              &                   &  \\
  FRDM~\cite{Moeller95}          & -1.01        &            &            &   0.25   &              & $ 1/2+$           &   \\
  \hline
  \end{tabular}
 \label{Table1}
 \end{table}

\newpage

%
%
\end{document}